\documentclass[11pt,psfig]{article}
\usepackage{graphics,graphicx}
\usepackage{enumerate}
\usepackage{amssymb,amsmath,amsthm,amsfonts,amssymb,mathtools}
\usepackage{xcolor,color}
\usepackage{bm}
\usepackage[numbers]{natbib}
\usepackage{hyperref}
\usepackage{academicons}
\usepackage{tikz}
\usepackage[ruled,vlined]{algorithm2e}
\usepackage[T1]{fontenc}
\usepackage{units}
\usepackage[normalem]{ulem}

\usepackage{array}

\definecolor{mygreen}{rgb}{0.1, 0.7, 0.3}
\usepackage{tcolorbox}
\usepackage{todonotes}

\newcommand{\ignore}[1]{}

\numberwithin{equation}{section}

\newcommand{\ket}[1]{|#1\rangle}
\newcommand{\bra}[1]{\langle#1|}
\newcommand{\Tr}{\text{Tr}}

\newcommand{\x}{\bm{x}}
\newcommand{\dd}{\bm{d}}
\newcommand{\e}{\bm{e}}
\newcommand{\s}{\bm{s}}
\newcommand{\y}{\bm{y}}
\newcommand{\ii}{\text{i}}

\newcommand{\R}{\mathbb R}
\newcommand{\C}{\mathbb C}

\newcommand{\Vt}{{\widetilde V}}
\newcommand{\Lt}{{\widetilde \Lambda}}
\newcommand{\Vh}{{\widehat V}}

\usepackage{sidecap,tikz}
\DeclareRobustCommand{\orcidicon}{\hspace{-1.0mm}
	\begin{tikzpicture}
		\draw[lime, fill=lime] (0.0,0.0) 
		circle [radius=0.15] 
		node[white] {{\fontfamily{qag}\selectfont \tiny \,ID}};
		\draw[white, fill=white] (-0.0525,0.095) 
		circle [radius=0.007];
	\end{tikzpicture}
	\hspace{-3.0mm}
}
\newcommand{\orcidDM}{\href{https://orcid.org/0000-0002-0166-4760}{\orcidicon}}
\newcommand{\orcidBH}{\href{https://orcid.org/0000-0002-8847-2058}{\orcidicon}}
\newcommand{\orcidVAM}{\href{https://orcid.org/0000-0002-9015-1234}{\orcidicon}}

\usepackage{mathtools}
\DeclarePairedDelimiterX\braket[2]{\langle}{\rangle}{#1\,\delimsize\vert\,\mathopen{}#2}

\graphicspath{{figs/}} 

\begin{document}
\date{\small\textsl{\today}}
\title{LREI: A fast numerical solver for quantum Landau–Lifshitz equations}
\author{
{\large Davoud Mirzaei\orcidDM}\thanks{Emial address: davoud.mirzaei@it.uu.se} \\
            Department of Information Technology, Division of Scientific Computing, \\ Uppsala University, 
            Box 337, SE-751 05 Uppsala, Sweden\\ \\
{\large Behnam Hashemi\orcidBH}\thanks{Email address: b.hashemi@le.ac.uk} \\ 
            School of Computing and Mathematical Sciences, University of Leicester, \\  Leicester, LE1 7RH, UK \\ \\
{\large Vahid Azimi-Mousolou\orcidVAM}\thanks{Email address: vahid.azimi-mousolou@physics.uu.se} \\ 
            Department of Physics and Astronomy, Divition of Material Theory, \\ Uppsala University, Box 516, 
            SE-751 20 Uppsala, Sweden 
 }

\maketitle
\begin{abstract}

We develop LREI, short for Low-Rank Eigenmode Integration, a memory- and time-efficient numerical scheme for solving quantum Landau–Lifshitz (q-LL) and quantum Landau–Lifshitz–Gilbert (q-LLG) equations, which govern spin dynamics in open quantum systems. Although the system size grows exponentially with the number of spins, our approach benefits from the low-rank structure of the density matrix and the sparsity of system Hamiltonians to avoid costly full matrix computations. By representing density matrices in terms of their low-rank factors and using Krylov subspace techniques for partial eigendecompositions, we reduce the per-step complexity of Runge–Kutta and Adams–Bashforth schemes from $\mathcal{O}(N^3)$ to $\mathcal{O}(r^2N)$, where $N = 2^n$ is the Hilbert space dimension for $n$ spins and $r \ll N$ is the effective rank of the density matrix. Likewise, the memory footprint is reduced from $\mathcal{O}(N^2)$ to $\mathcal{O}(rN)$, since no full $N\times N$ matrices are ever formed.
Among several technical improvements we applied, one key idea was to handle the computation of the action of the invariant subspace of the density matrix associated with its zero eigenvalues. This was accomplished by applying Householder reflectors constructed for the dominant eigenspace, thereby enabling the entire solution process to proceed without ever forming any large matrices. As an example, we can now evolve a time step of a twenty-spin system---corresponding to a density matrix size exceeding one million---in just a few seconds on a standard laptop. 
In addition, both classes of Runge-Kutta and Adams-Bashforth techniques are reformulated to preserve the physical properties of the density matrix throughout the time evolution. The new low-rank algorithm makes it possible to simulate much larger spin systems that were previously computationally infeasible. This, in turn, provides a powerful tool for  comparing the q-LL and q-LLG dynamics, assessing the validity of each model, and exploring how quantum features such as correlations and entanglement evolve across different regimes of system size and damping.
\end{abstract}

\section{Introduction}\label{sect:intro}
While the Landau–Lifshitz (LL) and Landau–Lifshitz–Gilbert (LLG) equations have long been central tools for modeling the microscopic dynamics of magnetic systems and materials, it is well understood that the underlying processes at the atomic scale are inherently quantum mechanical and governed by quantum dynamics. Crucially, intrinsically quantum features such as many-body {\em correlations} and {\em entanglement}, which are at the heart of modern quantum science and technology \cite{Amico2008, Chiara2018, Laflorencie2016, Laurell2025} lie beyond the descriptive power of classical equations. Thus, to obtain a more accurate and fundamental understanding of the dynamics of many-body spin systems, it is necessary to study quantum-mechanical counterparts of the LL and LLG equations.

A quantum analogue of the LL equation was introduced in \cite{wieser13} based on a phenomenological open-system framework. The motivation was to derive spin dynamics from a more fundamental quantum perspective than that offered by the classical LL equation. The q-LL equation reads as
  \begin{equation}\label{eq:q-LL}
\begin{split}
    &\dot{\rho} = \frac{\text{i}}{\hslash}[\rho, H] - \frac{\kappa}{\hslash} [\rho,[\rho,H]] \\
    &\rho(0) = \rho_0
    \end{split}
\end{equation}
  where $H$ is the Hamiltonian, $\rho\in \C^{N\times N}$ is the density operator\footnote{In line with standard conventions in numerical linear algebra, we denote matrices by capital letters. The only exceptions are the variable $\rho$ (with or without subscripts), and the Pauli matrices $\sigma_x$, $\sigma_y$, and $\sigma_z$ where we retain the standard notation from physics texts.}, 
$\dot{\rho} = \frac{d \rho }{dt}$, 
$\hslash$ is the Planck constant, $\kappa$ is the dimensionless damping rate, and $\rho_0$ is the initial condition. Here, the commutator, $[A,B]$, of two operators $A$ and $B$, is defined to be $[A,B]=AB-BA$.

  Building on this concept, in the recent paper \cite{liu24} a quantum analog of the classical Landau–Lifshitz–Gilbert (LLG) equation was presented as
\begin{equation}\label{eq:q-LLG}
\begin{split}
    &\dot{\rho} = \frac{\text{i}}{\hslash}[\rho, H] + \text{i}{\kappa} [\rho,\dot{\rho}] \\
    &\rho(0) = \rho_0
    \end{split}
\end{equation}
The q-LL model \eqref{eq:q-LL} has a simpler damping term compared with the q-LLG equation \eqref{eq:q-LLG}, since the time derivative $\dot\rho$ does not appear on its right-hand side. In fact, the q-LL equation is an approximation of the full model \eqref{eq:q-LLG}, obtained by substituting 
$\dot \rho$ on the right-hand side of \eqref{eq:q-LLG} with the whole right-hand side and truncating after the first term.

These models reveal several nontrivial quantum phenomena in systems composed of spin-$\frac{1}{2}$ particle pairs, including the emergence of spinless local states in antiferromagnetically coupled particles and 
emergence of long-lived highly entangled states, despite the presence of dissipation. 

However, the studies in \cite{wieser13,liu24} are limited to either pure state initial conditions or single- and two-spin systems, while quantum effects become increasingly intricate in general cases of initial conditions and/or many-spin settings. Although exact solutions exist for some special cases, a robust numerical solver is needed for studying the models in their general forms to further investigate their theoretical and practical aspects.

Two of the essential properties of the solution $\rho(t)$ of either equation \eqref{eq:q-LL} or equation \eqref{eq:q-LLG} are that it is Hermitian and possesses 
the conservation of the spectrum, i.e., 
\begin{equation}\label{eq:spectrum_conservation}
\frac{d}{dt} \lambda_k(\rho(t)) = 0
\end{equation}
where $\lambda_k\equiv \lambda_k(\rho(t))$, $k=1,\ldots,N$ are eigenvalues of $\rho(t)$. This means that the spectrum is independent of time; $\lambda_k(\rho(t)) = \lambda_k(\rho(0))$ for all $t\geq 0$. 
As a consequence, the traces of all powers of $\rho$ are {also} conserved;   
\begin{equation*}\label{eq:trace_conservation}
\frac{d}{dt} \text{Tr}(\rho^m) = 0, \quad m=1,2,\ldots  . 
\end{equation*}
The eigenvalues of $\rho(t)$ ($\equiv$ eigenvalues of $\rho_0$) are nonnegative, hence $\rho(t)$ remains positive semi-definite throughout the time integration. 

Exact solutions of \eqref{eq:q-LL} and \eqref{eq:q-LLG} exist only for pure (rank-one) initial states 
$\rho_0$. For the q-LL equation, the solution is
\begin{equation}\label{eq:exact_q-LL}
\rho(t) = \frac{\exp\left( -\frac{\ii}{\hslash}\tilde Ht\right)\rho_0 \exp\left(\frac{\ii}{\hslash}\tilde Ht\right)}{\mathrm{Tr}\left( \exp\left( -\frac{\ii}{\hslash}\tilde Ht\right)\rho_0 \exp\left(\frac{\ii}{\hslash}\tilde Ht\right)\right)},
\end{equation}
where $\tilde H = \left(1 - \ii\kappa \right)H$. The exact solution of the q-LLG equation in the pure-state case coincides with \eqref{eq:exact_q-LL}, up to a rescaling of time $t\mapsto t/(1+\kappa^2)$. We use these closed-form solutions to assess the accuracy of our proposed algorithm in such special cases. However, the exact solutions involve computing matrix exponentials of the form $\exp(A)$ for complex matrices $A$, which in turn requires robust and efficient numerical solvers.

Any numerical method for solving the q-LL and q-LLG equations in their general form should ideally be robust, highly accurate, preserve the physical properties of the models, and scale to large spin systems. Numerical simulation becomes particularly challenging in many-body settings due to the exponential growth of the density matrix $\rho$, which is a {\em full} complex matrix of size $N \times N$ with $N = 2^n$, where $n$ is the number of particles. This exponential scaling, combined with the inherent nonlinearity of the equations, poses severe computational challenges for simulating large quantum systems. For related numerical approaches to simulating large quantum systems, we refer the reader to \cite{CaiLu2018, WangCai2022}.

In \cite{azimi-mirzaei:2025}, we proposed a numerical method for solving the q-LLG equation \eqref{eq:q-LLG}, which can be readily adapted to the q-LL equation \eqref{eq:q-LL}. The method efficiently handles the complexities of quantum spin systems while preserving the key physical properties of the q-LLG dynamics. While it represents the first numerical solution of this equation, the approach relies on eigendecomposition of the evolving density matrix at each time step. As a result, the method becomes computationally expensive for large-scale systems due to the exponential growth of the matrix sizes with increasing numbers of spins.

In this paper we exploit the observation that in many practical scenarios the matrices and operators involved are initially of {\em low rank} and retain this property throughout the time evolution. By combining low-rank matrix representations with {\em Krylov subspace methods} and structure-preserving time integration, we develop an efficient computational method which reduces the memory usage and the time complexity, significantly. This enables the simulation and study of much larger quantum spin networks. 

The remainder of the paper is organized as follows. In Section \ref{sect:oldmethod}, we briefly review the method of \cite{azimi-mirzaei:2025}. Section \ref{sect:lrei} introduces our new method for solving the q-LLG equation \eqref{eq:q-LLG}, and Section \ref{sect:q-LL} demonstrates how this method can be adapted to the q-LL equation \eqref{eq:q-LL}. In fact, we first address the more complex q-LLG equation and then tailor the approach to the simpler q-LL model. Interestingly, from a computational perspective, both equations exhibit essentially the same complexity. Section \ref{sect:numerical} presents numerical experiments, and Section \ref{sect:summury} concludes with a summary and directions for future research.

\section{Available numerical solution}\label{sect:oldmethod}

In \cite{azimi-mirzaei:2025}, a conservative modification of explicit Runge-Kutta methods was developed for solving the q-LLG equation \eqref{eq:q-LLG}. The approach is briefly outlined below.

We starting with the eigendecomposition
\begin{equation}\label{eq:eigendecom}
  \rho = V\Lambda V^* 
\end{equation}
where $V\in\C^{N\times N}$ is a unitary matrix whose columns are the eigenvectors of $\rho$, and $\Lambda\in \R^{N\times N}$ is a diagonal matrix containing the eigenvalues\footnote{Here $V^*$ denotes the Hermitian conjugate (adjoint) of $V$. This convention differs from that in standard quantum information texts such as \cite{nielsen-et-al:2002}, where the star symbol refers only to the element-wise complex conjugate of a matrix.}. The change of variables 
\begin{equation}\label{eq:Xdef}
X := V^*\dot{\rho}\, V,
\end{equation}
transforms the original equation into a {\em Sylvester-type equation} of the form 
\begin{equation}\label{eq:q-LLG-3}
(I + S)X - XS = D,
\end{equation}
where 
\begin{equation}\label{eq:Dmat}
    D = \frac{\ii}{\hslash} V^*[\rho, H]V
\end{equation}
depends nonlinearly on $\rho$ and $S = -\ii \kappa \Lambda$. 
The matrix $X$ is computed from \eqref{eq:q-LLG-3} and used to evolve $\rho$ via
\begin{equation}\label{eq:q-LLG-final}
\begin{split}
&\dot{\rho} = V X V^*, \\
&\rho(0) = \rho_0,
\end{split}
\end{equation}
instead of the original form \eqref{eq:q-LLG}. Notably, the right-hand side $V X V^*$ is now independent of $\dot{\rho}$ and depends only on $\rho$ through its spectral decomposition.

The solution to the Sylvester-type equation \eqref{eq:q-LLG-3} is given column-wise as
\begin{equation}\label{eq:x_ell_sol}
\x_\ell = \text{cwd}(\dd_\ell, \e + \s - s_\ell \e), \quad \ell = 1, 2, \ldots, N,
\end{equation}
where $\x_\ell$ and $\dd_\ell$ denote the $\ell$-th columns of $X$ and $D$, respectively, $\s$ is the diagonal of $S$; and $\e = [1, 1, \ldots, 1]^T$. The operator $\text{cwd}(\x, \y)$ denotes the componentwise division of vectors $\x$ and $\y$. The solution is well-defined because the matrices $I + S$ and $-S$ have no eigenvalues in common, which, in turn, follows from the fact that $\rho$, being Hermitian, has no eigenvalues equal to ${-\ii}/{2\kappa}$.

Explicit Runge-Kutta (RK) methods were then employed to integrate equation \eqref{eq:q-LLG-final} in time, as described in \cite{azimi-mirzaei:2025}. To preserve the spectrum of the density matrix during the time integration, the approach in \cite{azimi-mirzaei:2025} projects all the intermediate RK states back onto the manifold of matrices that share the same spectrum as the initial density matrix $\rho_0$ by replacing the evolving eigenvalue matrix $\Lambda(t)$ with the fixed initial eigenvalue matrix $\Lambda(0)$ at each stage of the RK method.
We refer to such techniques as the {\em eigenmode integration (EI) methods} for brevity. For instance, when combined with RK schemes, we refer to them as EI-RK methods. 

\subsection{Computational bottleneck}
The dominant computational cost of {EI methods} comes from the repeated eigendecompositions of the full density matrix $\rho\in \C^{N\times N}$ at each time step. Specifically, for an $s$-stage Runge-Kutta method, $s$ eigendecompositions are required per time step. Given that $N = 2^n$, where $n$ is the number of spins, and each decomposition costs $\mathcal O(N^3)$, the total cost at each time step grows exponentially with $n$, more precisely  $\mathcal{O}(s N^3)$. As a result, both the {\em time} and {\em memory} complexities of the algorithm increase significantly as the system size grows. This is a major computational bottleneck for large spin systems. The aim of this paper is to benefit from low-rank structure of the density matrix $\rho$ and the sparsity of other matrices involved to reduce the time complexity to $\mathcal{O}(s r^2 N)$, and the space complexity to $\mathcal{O}(s r N)$, where $r\ll N$ is the rank of $\rho$. 

\section{Low-rank eigenmode integration (LREI) scheme for q-LLG}\label{sect:lrei}

In typical physical systems, such as spin chains, lattices, and quantum many-body models, the Hamiltonian $H$ is a sparse matrix due to local interactions and is generally full-rank in the absence of special symmetries or fine-tuning.
Moreover, in simulations and experiments—particularly in quantum dynamics—the initial density matrix $\rho_0$ is often low-rank, either inherently or as a result of renormalization or truncation, especially at low temperatures or when representing only a small subset of relevant eigenstates. We assume that
$$
\operatorname{rank}(\rho_0) = r
$$ 
with $r \ll N$. Owing to the structure-preserving nature of the q-LLG evolution, the rank of $\rho(t)$ remains unchanged and satisfies 
$$
\operatorname{rank}(\rho(t))  = r, \quad t\geq 0.
$$ 
We benefit from the sparsity of $H$ and mostly the low-rank structure of $\rho(t)$ to develop an efficient algorithm, which we refer to as the {\em low-rank eigenmode integration (LREI)} method. In this section, we first develop the LREI method for solving the q-LLG equation and then adapt it to the q-LL equation in Section \ref{sect:q-LL}.

\subsection{Partial eigendecomposition}
The eigendecomposition \eqref{eq:eigendecom} can be rewritten as
\begin{equation*}
    \rho = \begin{bmatrix} \Vt & \Vh \end{bmatrix}
           \begin{bmatrix} \Lt &  \\ & \bm{0} \end{bmatrix}
           \begin{bmatrix} \Vt^* \\ \Vh^* \end{bmatrix}
\end{equation*}
where $\Vt \in \C^{N \times r}$ contains orthonormal columns corresponding to the $r$ positive eigenvalues of $\rho$, and $\Lt \in \R^{r \times r}$ is a diagonal matrix with these eigenvalues sorted in a decreasing order on its diagonal. In the quantum computing language, $\Vt$ is called the {\em support} of operator $\rho$.
The partial eigendecomposition 
\begin{equation}\label{eq:partial-eigendecom}
\rho = \Vt \Lt \Vt^*,
\end{equation}
can be computed efficiently using Krylov subspace iterative methods, which are well-suited for large-scale Hermitian matrices; see Section~\ref{sect:rk} for more details.

However, the contribution of the complementary subspace, associated with the matrix $\Vh\in \C^{N \times (N-r)}$ 
also plays a significant role in the numerical procedure described above. In Section \ref{sect:AV}, by relying only on partial decomposition 
\eqref{eq:partial-eigendecom}, 
 we will effectively handle the contribution of big matrix $\Vh$ without explicitly forming and storing it.  

\subsection{Space complexity}
To manage the space (memory) complexity of the algorithm, we avoid storing or manipulating any {\em full} matrices of size $N \times N$ or $N \times (N - r)$, since $N = 2^n$ grows exponentially with the number of spins $n$. This applies in particular to matrices such as $\rho$, $\Vh$, $D$, and $X$.

According to the underlying theory, the spectrum $\Lambda$ of $\rho$ remains invariant throughout the time integration. To enforce this property in our numerical scheme, we adopt the same projection method proposed in \cite{azimi-mirzaei:2025}. As a result, instead of explicitly working with $\rho$, we store and evolve only its factor $\Vt$ of size $N\times r$  throughout the computation. Furthermore, in the remainder of this section, we explain in detail how to completely avoid manipulating the large matrices $\Vh$, $D$, and $X$.

\subsection{Efficient computation of $X$}

Equations \eqref{eq:x_ell_sol} can be written in a matrix form as 
\begin{equation*}
    X = \text{cwd}(D, E +  L - L^T)
\end{equation*}
where $D$ is defined in \eqref{eq:Dmat}, $E$ denotes the $N\times N$ matrix of all ones, and $ L = -\ii\kappa [\bm{\lambda} \; \bm{\lambda} \;\ldots\; \bm{\lambda} ]\in \C^{N\times N}$, where $\bm{\lambda}= [\lambda_1,\ldots,\lambda_r,0,\ldots,0]^T$ is the vector of all eigenvalues of $\rho$. Here, the operator $\text{cwd}(A, B)$ denotes the componentwise division of matrices $A$ and $B$. We also note that according to the structure of $L$ we simply have $-L^T = L^*$. On the other hand, we can split the matrix $L$ into four blocks as
$$
L = \begin{bmatrix}
    L_{11} & L_{12} \\ \bm{0} & \bm{0}
\end{bmatrix}
$$
where $L_{11}$ is an $r\times r$ matrix and $L_{12}$ is of size $r\times (N-r)$. If we split the all-ones matrix $E$ in the same way, we can write the denominator as
\begin{equation*}
    E +  L - L^T = E + L+L^* = \begin{bmatrix}
    E_{11}+L_{11}+L_{11}^* & E_{12}+L_{12} \\ (E_{12}+L_{12})^* & E_{22}
\end{bmatrix} .
\end{equation*}
In addition, from \eqref{eq:Dmat} the numerator
$D$ has the representation 
\begin{align*}
   D &= \frac{\ii}{\hslash} V^*[\rho, H]V = \frac{\ii}{\hslash}\begin{bmatrix}
     \Vt^*\\ \Vh^* 
  \end{bmatrix}  \begin{bmatrix}
      \Vt\Lt \Vt^* & H
  \end{bmatrix}
  \begin{bmatrix}
      \Vt & \Vh
  \end{bmatrix} \\
  & = \frac{\ii}{\hslash}\begin{bmatrix}
      \Lt\Vt^*H\Vt-\Vt^*H\Vt\Lt & \Lt\Vt^*H\Vh\\ 
      -\Vh^*H\Vt\Lt & \bm{0}
  \end{bmatrix} \\
  &=:  \begin{bmatrix}
      D_{11} & D_{12}\\ 
      D_{12}^* & \bm{0}
  \end{bmatrix}   
\end{align*}
where the {top-left} block $D_{11}$ is of size $r\times r$ and $D_{12}$ is of size $r\times (N-r)$. This shows that the major part of $D$ (the fourth block) is zero. In the computation above we used the facts that $\Vt^*\Vt=I$, $\Vt^*\Vh = \bm{0}$, $\Vh^*\Vt = \bm{0}$, and $H=H^*$. 

Consequently, the solution $X$ can be similarly split into four blocks as
$$
X = \begin{bmatrix}
    X_{11} & X_{12} \\ X_{12}^* & \bm{0}
\end{bmatrix}
$$
where 
\begin{equation}\label{eq:X-blocks-sol}  
\begin{split}
   & X_{11} = \text{cwd}(D_{11}, E_{11}+L_{11}+L_{11}^*) , \quad D_{11} = \frac{\ii}{\hslash} (\Lt\Vt^*H\Vt-\Vt^*H\Vt\Lt),\\
   & X_{12} = \text{cwd}(D_{12}, E_{12}+L_{12}), \quad D_{12} = \frac{\ii}{\hslash}\Lt\Vt^*H\Vh. 
\end{split}
\end{equation}

The computation of $X_{11}$ is relatively inexpensive due to the sparsity of the large matrix $H$, the small size of $\Lt$, $L_{11}$ and $E_{11}$ (all $r \times r$), and the tall-skinny structure of $\Vt$. 

The main computational challenge lies in computing $X_{12}$, as it involves the dense and high-dimensional matrix $\Vh$ through the computation of $D_{12}$. To address this, we avoid explicitly constructing $\Vh$ and instead employ an efficient technique for computing $D_{12}$ implicitly, as detailed below.

\subsection{Efficient computation of products $A\Vh$ and $\Vh \! A$}\label{sect:AV}

In this section, we describe how to efficiently compute matrix products of the form $A\Vh$, which are required for evaluating $D_{12}$, and $\Vh\! A$ which will later be needed for assembling the right-hand side of the ODE \eqref{eq:q-LLG-final}.
Recall that the matrices $\Vt$ and $\Vh$ contain eigenvectors corresponding to the positive and zero eigenvalues, respectively, of the positive semi-definite matrix $\rho$. As such, the columns of $\Vh$ form the orthonormal complement of the columns of $\Vt$. 
One common approach to computing $\Vh$, without having it at hand, is via the full QR factorization of the $N \times r$ matrix $\Vt$:
$$
\Vt = QR = \begin{bmatrix} \widetilde Q & \widehat Q \end{bmatrix}
\begin{bmatrix}
\widetilde R \\
\bm{0}
\end{bmatrix},
$$
where $Q$ is a unitary matrix with $\widetilde Q \in \C^{N \times r}$, $\widehat Q \in \C^{N \times (N - r)}$, and $\widetilde R \in \C^{r \times r}$ is upper triangular. By normalizing the diagonal entries of $\widetilde R$ to be real positive numbers, we can ensure that $\widetilde Q = \Vt$ (up to a column permutation), and hence $\widehat Q$ provides an orthonormal basis for the orthogonal complement, i.e., $\Vh = \widehat Q$. The full unitary matrix $Q = [\widetilde Q \;\; \widehat Q]$ therefore coincides with the complete eigenvector matrix $V$.

The QR factorization of $\Vt$ {via Householder algorithm} involves $r$ steps, each using a {\em Householder transformation}
\begin{equation}\label{eq:Householder-mat}
P_k = I - \frac{2}{\bm{u}_k^* \bm{u}_k} \bm{u}_k \bm{u}_k^*, \quad k = 1, \ldots, r,
\end{equation}
where each $P_k$ is Hermitian and unitary \cite{datta:2010, TB:2022}. These matrices are applied sequentially to introduce zeros below the diagonal in the $k$-th column of $\Vt$, eventually producing the upper triangular matrix $R$. The full matrix $Q = V$ is then expressed as the product of Householder reflectors, i.e.,
$$
Q = V = P_1 P_2 \cdots P_r.
$$
To compute matrix products such as $AV$ or $V\! A$ without explicitly forming the large matrix $V$, we can exploit the special structure of the Householder matrices. Instead of constructing each $P_k$ in full, we work directly with the {Householder} vectors $\bm{u}_k$ that define them via \eqref{eq:Householder-mat} to have an efficient and memory-saving implementations of such matrix-matrix products.

More precisely, instead of performing a full QR factorization, we compute only the $r$ Householder vectors $\bm{u}_k$, for $k = 1, 2, \ldots, r$. This requires approximately $2Nr^2$ floating-point operations (flops), which is computationally cheap when $r$ is small.
To compute matrix products of the form $AV = AP_1P_2 \cdots P_r$, one would naively perform $r$ successive matrix-matrix multiplications of the form $AP$. For a general $m \times N$ matrix $A$ and a dense $N \times N$ matrix $P$, this would cost approximately $2mN^2$ flops. However, by using the structure of the Householder matrix
$$
P = I - \beta \bm{u}\bm{u}^*, \quad \beta = \frac{2}{\bm{u}^*\bm{u}},
$$
we can compute the product efficiently without forming $P$ explicitly. We use the identity
$$
AP = A(I - \beta \bm{u}\bm{u}^*) = A - \beta \bm{w} \bm{u}^*, \quad \bm{w} = A \bm{u},
$$
which requires only $4mN$ flops.
A similar strategy applies for computing $V\! A = P_1P_2 \cdots P_r A$ for a matrix $A \in \C^{N \times m}$. Using the same formulation,
$$
PA = (I - \beta \bm{u}\bm{u}^*)A = A - \beta \bm{u} \bm{w}^*, \quad \bm{w} = A^* \bm{u},
$$
also costs for $4mN$ flops. In our algorithm, $m = r$, and since this product must be repeated $r$ times, the total cost of computing $AV$ or $V\! A$ becomes $4Nr^2$.
However, we are only interested in the portion involving $\Vh$, the orthogonal complement of $\Vt$. For computing $A\Vh$, observe that
$$
AV = A[\Vt \;\; \Vh] = [A\Vt \;\; A\Vh],
$$
so we compute $AV$ and extract only the last $N - r$ columns corresponding to $A\Vh$.

To compute $\Vh\! A$, where $A \in \C^{(N - r) \times r}$, we augment $A$ by placing a zero block of size $r \times r$ on top and obtain
$$
 \Vh A = [\Vt \;\; \Vh] \begin{bmatrix} \bm{0} \\ A \end{bmatrix} =: V\! A_0.
$$
Then the product can be computed using the same Householder-based approach discussed above.

Consequently, the total cost of computing the products $A\Vh$ for $A \in \C^{r \times N}$, and $\Vh\! A$ for $A \in \C^{r \times (N-r)}$, is approximately $6Nr^2$. Since $r$, the rank of $\rho$, is fixed and typically small, the overall time complexity becomes $\mathcal{O}(N)$, which is optimal in terms of $N$, the size of the system.
Importantly, throughout this process, we never form the large matrix $\Vh$ or the full Householder matrices $P_k$ explicitly. Instead, we operate directly on the small portion $\Vt$ and the compact Householder vectors $\bm{u}_k$, to reduce both memory and computational costs, significantly. The space complexity is clearly of order $rN$. 

\subsection{Computation of the ODE right-hand side $VXV^*$}

All ODE solvers used to integrate \eqref{eq:q-LLG-final} require efficient evaluation of the right-hand side matrix $V X V^*$ at each iteration. Although this matrix is of size $N \times N$, we avoid forming it explicitly by working only with its low-rank ingredients. Suppose that, at a given time step, we are provided with the partial factor $\Vt$ and positive part of the spectrum, $\Lt$, of $\rho$. As described in the previous subsections, this information suffices to compute the submatrices $X_{11} \in \C^{r \times r}$ and $X_{12} \in \C^{r \times (N - r)}$ of the matrix $X$, with time and memory complexities of  $\mathcal{O}(r^2N)$ and $\mathcal{O}(rN)$, respectively. 
Now, employing the block representations, we can write
\begin{align*}
    VXV^* &= [\Vt \;\; \Vh] \begin{bmatrix}
        X_{11} & X_{12}\\ X_{12}^* & \bm 0
    \end{bmatrix}\begin{bmatrix}
        \Vt^* \\ \Vh^*
    \end{bmatrix} \\
    &= \Vt X_{11}\Vt^* +  \Vh X_{12}^*\Vt^* + \Vt X_{12}\Vh^* \\
    &= (\Vt X_{11} + \Vh X_{12}^*)\Vt^* + \Vt(X_{12}\Vh^*) \\
    & = Z\Vt^* + \Vt W^*
\end{align*}
where 
\begin{align*}
    Z = \Vt X_{11} + W, \quad W = \Vh X_{12}^*. 
\end{align*}
The computation of $W$ involves the orthogonal complement matrix $\Vh$, but as shown in the previous subsection, we avoid constructing $\Vh$ explicitly. Instead, we compute $W$ efficiently using the Householder-based approach with a cost of $\mathcal{O}(Nr^2)$. The first term $\Vt X_{11}$ of $Z$ is inexpensive to compute since $X_{11}$ is a small $r \times r$ matrix.

Thus, rather than forming the full $N \times N$ matrix $V X V^*$, we return the pair of matrices $Z \in \C^{N \times r}$ and $W \in \C^{N \times r}$, and express the time derivative as
\begin{equation}\label{eq:q-LLG-low-rank}  
\dot{\rho} = Z \Vt^* + \Vt W^*.
\end{equation}
We indeed write the right-hand side of the ODE in terms of its low-rank components $\Vt$, $Z$, and $W$. 
We will see in Subsections \ref{sect:rk} and \ref{sect:ab}, how this specific representation can be efficiently used in numerical ODE solvers. 

\subsection{Memory-efficient representation of $\rho(t)$}\label{sect:rho-computation}

Since the spectrum of $\rho(t)$ is preserved over time, we have
$$
\rho(t) = \Vt(t) \Lt(0) \Vt(t)^*,
$$
where $\Lt(0)=:\Lt_0$ denotes the fixed diagonal matrix of positive eigenvalues of $\rho(0)$. This allows us to avoid storing or evolving the full $N \times N$ matrix $\rho(t)$. Instead, we restrict the evolution to the low-rank factor 
$\Vt(t)$, which yields substantial dimensionality reduction without loss of any information. In what follows, we show how physical observables such as {\em energy}, {\em magnetization}, and {\em entanglement measures} can be computed efficiently using only $\Vt(t)$, without forming $\rho(t)$ explicitly.

For the energy and magnetization, we compute the corresponding expectation values 
\begin{equation*}
\langle A\rangle=\Tr[A\rho(t)]
\end{equation*}
where $A$ represents either the Hamiltonian $H$ (for energy), or the magnetization operators $M_v$, for $v = x, y, z$ (for magnetization). The magnetization operator along direction $v$ is defined by
$$
M_v = \frac{1}{n} \sum_{i=1}^n S_i^v,
$$
with $S_i^v$ denoting the spin operator at site $i$ in the $v$-direction,
\begin{equation}\label{eq:spin_op}
S_i^v = \underbrace{I\otimes I \otimes \cdots \otimes I}_{(i-1) \text{ times}} \otimes \frac{\hslash}{2} \sigma_v\otimes \underbrace{I \otimes \cdots \otimes I}_{n-i \text{ times}}  , \quad v = x,y,z,
\end{equation}
where $I$ is the identity matrix, and $\sigma_x, \sigma_y, \sigma_z \in \mathrm{SU}(2)$ are the Pauli matrices \cite{Nielsen2010}.

Although both $H$ and $M_v$ are large $N \times N$ matrices, their sparse structure allows them to be stored in memory-efficient formats. Using the factorized form of $\rho(t)$, the expectation value becomes
\begin{equation}\label{eq:measuerA}
\langle A\rangle = \Tr[A\rho(t)] = \Tr[A\Vt\Lt_0\Vt^*] = \Tr[\Lt_0(\Vt^*(A\Vt))]
\end{equation}
where the last equality follows from the cyclic property of the trace. The parentheses are deliberately placed to enforce an optimized order of matrix multiplication. We first multiply the sparse matrix $A$ by the tall-skinny matrix $\Vt$, yielding an $N \times r$ matrix. This result is then left-multiplied by $\Vt^*$ to produce a small $r \times r$ matrix, which is finally multiplied by the diagonal matrix $\Lt_0$ of size $r \times r$. This ordering minimizes both memory usage and computational cost by replacing the full trace computation with one of complexity $\mathcal{O}(Nr^2)$, where $r \ll N$. 

To compute the entanglement between two spins located at sites $k$ and $l$, the measurement is no longer linear with respect to the density operator and cannot be directly expressed in the simple form of an expectation value as in \eqref{eq:measuerA}. In fact, two-spin entanglement can be quantified by measures such as concurrence \cite{Wootters1998}, negativity \cite{Vidal2002}, or nonlocality \cite{Horodecki1995}, all of which involve non-linear operations on the reduced two-spin density matrix $R_{kl}(t) \in \mathbb{C}^{4 \times 4}$ of the global state $\rho(t)$. This reduced density matrix is obtained via the partial trace
$$
R_{kl}(t) = \Tr_{\{\mathbf{S}_i\}_{i=1\,\&\,i\ne k,l}^{n}}\, \rho(t),
$$
where the trace is taken over all degrees of freedom (i.e., basis states) corresponding to spins not in $\{\mathbf{S}_k, \mathbf{S}_l\}$, with each $\mathbf{S}_i = [S_i^x, S_i^y, S_i^z]$ denoting the spin operator vector at site $i$ \cite{Nielsen2010}\footnote{A boldface capital letter, such as $\mathbf{S}$, denotes a horizontal concatenation of three matrices.}.
The computation of the 16 entries of the matrix $R_{kl}$ reduces to evaluating expressions of the form $\Tr(A \rho(t))$ for certain sparse observables $A$, similar to the energy and magnetization computations described earlier. Subsequent entanglement measures (such as concurrence, negativity, and locality) are then computed directly from the small $4 \times 4$ matrix $R_{kl}(t)$.

\subsection{Explicit one-step schemes}\label{sect:rk}

Explicit one-step ODE solvers start with the initial density matrix $\rho_0$  
and compute an approximate solution $\rho_{k+1}\approx \rho(t_{k+1})$ at each time level $t_{k+1} = (k+1)h$ using a recurrence of the form
$$
\rho_{k+1} = \rho_k + h \psi(\rho_k;h), \quad k = 0, 1, 2, \ldots,
$$
where $h=t_{k+1}-t_k$ denotes the time step and $\psi$ is a function of $\rho_k$ and $h$ depending on the right-hand side of the ODE. The iteration continues until the final time $t_F$ is reached. For example, in Euler's method 
 $\psi(\rho;h) = Z\Vt^* + \Vt W^* =: f(\rho)$ and for a Runge-Kutta method of order two (RK2) $\psi(\rho;h) = \frac{1}{2}[f(\rho)+f(\rho+hf(\rho))]$ which can be reformulated as a two-stage scheme 
 $$
 \begin{array}{rl}
    K_1& = \rho_k\\
    K_2& = K_1+hf(K_1)\\ \hline 
  \rho_{k+1}& = K_1 + \frac{1}{2}h[f(K_1) + f(K_2)]
 \end{array}
 $$
where the function $f(K)$ is given by
$$
f(K) = Z \Vt^* + \Vt W^*,
$$
with $\Vt$ being the partial eigenvector matrix of $K$ corresponding to its nonzero eigenvalues. The matrices $Z$ and $W$ are computed as described previously, by evaluating the right-hand side of the evolution equation with $\rho$ replaced by $K$.
  
In general, an explicit $s$-stage RK method has the form
\begin{equation}\label{rk:sstageform}
\begin{array}{rl}
\displaystyle K_1 &=\, \rho_k\\
\displaystyle K_2 &=\, K_1 + h a_{2,1} f(K_1) \\
\displaystyle K_3 &=\, K_1 + h \left[ a_{3,1} f(K_1) + a_{3,2} f(K_2) \right] \\
&\vdots \\
\displaystyle K_s &=\, K_1 + h \left[ a_{s,1} f(K_1) + \cdots + a_{s,s-1} f(K_{s-1}) \right] \\  \hline 
\displaystyle \rho_{k+1} &=\, K_1 + h \left[ b_1 f(K_1) + b_2 f(K_2) + \cdots + b_s f(K_s) \right]
\end{array}
\end{equation}
for some real coefficients $a_{j,\ell}$ and $b_\ell$. For example, in the above RK2 scheme, $a_{2,1}=1$ and $b_1=b_2 = \frac{1}{2}$. We refer the reader to \cite{Butcher:2016-1} for details and the list of coefficients for higher order RK methods. At each stage, we compute $f(K_j) = Z_j\Vt_j^*+\Vt_jW_j^*$ and update the stage to obtain the new matrix $K_{j+1}$ via 
\begin{equation}\label{eq:K-expand}
\begin{split}
K_{j+1} &= K_1 + h \left[ a_{j+1,1} f(K_1) + \cdots + a_{j+1,j} f(K_j) \right] \\
&= \Vt_1\Lt_0\Vt_1^* + h\left[ a_{j+1,1} (Z_1\Vt_1^*+\Vt_1W_1^*) + \cdots + a_{j+1,j} (Z_j\Vt_j^*+\Vt_jW_j^*)
\right]\\
& =: A_1\Vt_1^*+ \Vt_1 {B_1^\ast} + A_2\Vt_2^*+ \Vt_2 {B_2^\ast} + \cdots + A_j\Vt_j^*+ \Vt_j {B_j^\ast}
\end{split}
\end{equation}
where 
$$
\begin{array}{ll}
   A_1 = \Vt_1\Lt_0 + ha_{j+1,1}Z_1, & B_1 = ha_{j+1,1}{W_1},\\
   A_\ell = ha_{j+1,\ell}\,Z_\ell,         & B_\ell = ha_{j+1,\ell} {W_\ell}, \quad \ell=2,\ldots,j,
\end{array}
$$
and finally 
\begin{equation}\label{eq:rho-expand}
\begin{split}
\rho_{k+1} &= K_1 + h \left[ b_{1} f(K_1) + \cdots + b_s f(K_s) \right] \\
&= \Vt_1\Lt_0\Vt_1^* + h\left[ b_1 (Z_1\Vt_1^*+\Vt_1W_1^*) + \cdots + b_s (Z_s\Vt_s^*+\Vt_s W_s^*)
\right]\\
& =: A_1\Vt_1^*+ \Vt_1 {B_1^\ast} + A_2\Vt_2^*+ \Vt_2 {B_2^\ast} + \cdots + A_s\Vt_s^*+ \Vt_s {B_s^\ast}
\end{split}
\end{equation}
where the new matrices $A_\ell$ and $B_\ell$ are defined by
$$
\begin{array}{ll}
   A_1 = \Vt_1\Lt_0 + hb_{1}Z_1, & B_1 = hb_1 {W_1}, \\
   A_\ell = hb_\ell Z_\ell,         & B_\ell = hb_\ell {W_\ell},\quad \ell=2,\ldots,s.
\end{array}
$$
We first point out that this scheme preserves the spectrum of $\rho_k$ for all $k = 0, 1, \ldots$, since the initial eigenvalue matrix $\Lt_0$ is consistently used in place of $\Lt_1$ in the second row of equations \eqref{eq:K-expand} and \eqref{eq:rho-expand}. This guarantees that the eigenvalues of $\rho_k$ remain unchanged throughout the evolution.

We also do not store the full matrices $K_j$ and $\rho_{k+1}$ as dense $N \times N$ arrays. We also do not store the full matrices $K_j$ and $\rho_{k+1}$ as dense $N \times N$ arrays. Instead, we use them in their low-rank representations (not approximation!) as in equations \eqref{eq:K-expand} and \eqref{eq:rho-expand}, involving matrices $A_\ell$, $B_\ell$, and $\Vt_\ell$, each of size $N \times r$. Note that $\rho_{k+1}$ computed at time step $k$ serves as $K_1$ for the next time step $k+1$. 

Given the sparse initial density matrix $\rho_0$ and its rank $r = \operatorname{rank}(\rho_0)$, we compute its partial eigendecomposition 
$$
\rho_0 = \Vt\Lt_0\Vt^*, \quad \Lt_0 = \Lt(0)
$$
and start the RK iterations. Each iteration $k$ contains $s$ stages. 
At each stage $j+1$, we must compute the partial eigendecomposition of the matrix $K_{j+1}$. This is performed efficiently 
using iterative {\em Krylov subspace methods}, more specifically the Implicitly Restarted Lanczos Method (IRLM) \cite{Lehoucq-et-al:1998, Stewart02}.

The idea is to build a length-$m$ Lanczos decomposition 
$$
K_{j+1} U_m = U_m T_m + \bm{f}_m \bm{e}_m^\ast
$$ 
where $m$ is the dimension of the Krylov subspace 
$$
\mathcal{K}_m(K_{j+1}, \bm{u}_1) := \{ \bm{u}_1, K_{j+1} \bm{u}_1, K_{j+1}^2 \bm{u}_1, \dots, K_{j+1}^{m-1} \bm{u}_1\}
$$
to project onto, $U_m$ is an $N \times m$ matrix whose columns form an orthogonal basis for the Krylov subspace, $e_m$ denotes the $m$-th column of the $m \times m$ identity matrix, $\bm f_m$ is the residual vector which is orthogonal to the columns of $U_m$, and $T_m = U_m^\ast K_{j+1} U_m$ is tridiagonal. The eigenvalues of the small $m \times m$ matrix $T_m$, known as the Ritz values, approximate those of $K_{j+1}$, and its eigenvectors when multiplied by $U_m$, yield the corresponding approximate eigenvectors of $K_{j+1}$.

Two points merit highlighting in practice. First, to obtain $r$ eigenvalues of $K_{j+1}$ to high-accuracy, it is enough to project onto a Krylov subspace with dimension as small as $m = 2r+1$, which is far less than $N$. Second, the method does not necessarily require the full matrix $K_{j+1}$ to be explicitly formed \cite{Lehoucq-et-al:1998, Stewart02}. Instead, these methods only require access to a function that computes matrix-vector products $K_{j+1} \x$ for arbitrary vectors $\x \in \C^N$.
This allows us to supply the Krylov solver with a function $g: \C^N \to \C^N$, defined by
\begin{equation}\label{eq:g-func}
g(\x) = A_1(\Vt_1^* \x) + \Vt_1 ({B_1^\ast} \x) + \cdots + A_j(\Vt_j^* \x) + \Vt_j ({B_j^\ast} \x),
\end{equation}
where all matrix-vector multiplications are performed using only the low-rank components. The parentheses in the expression for $g(\x)$ are deliberately placed to avoid forming full $N \times N$ matrices, instead, allowing only matrix-vector products involving matrices of size $N \times r$.  There are $m$ Lanczos steps, each involving matrix-vector products whose cost is $\mathcal{O}(Nr)$, summing to $\mathcal{O}(mNr) = \mathcal{O}(Nr^2)$ operations which is similar to the cost of orthonormalization. The resulting $m \times m$ tridiagonal eigenvalue problem requires $\mathcal{O}(m^2) = \mathcal{O}(r^2)$ operations for the Ritz values which is negligible for $r\ll N$ and forming $r$ Ritz vectors adds $\mathcal{O}(Nr^2)$. Overall, the time complexity of each execution of the Lanczos method remains $\mathcal{O}(Nr^2)$ and its storage is $\mathcal{O}(Nm) = \mathcal{O}(Nr)$.

Similarly, we avoid forming the full matrix $\rho_{k+1}$ and instead work directly with its low-rank representation \eqref{eq:rho-expand}, just as we did for the intermediate stages $K_{j+1}$.
From this representation, we compute its partial eigenvector matrix $\Vt_1$, which serves as the representative of $K_1$ matrix for the next time step. This iteration is repeated until the final time $t_F$ is reached.

We refer to this class of methods as LREI-RK schemes to highlight their foundation in low-rank eigendecomposition combined with Runge-Kutta time integration.

\subsection{Explicit multi-step schemes}\label{sect:ab}
Runge–Kutta methods belong to the class of single-step schemes, where at each time step $k+1$, the solution $\rho_{k+1}$ depends solely on the previous value $\rho_{k}$. However, each RK step involves $s$ intermediate {\em stages}, each requiring one partial eigendecomposition.

In contrast, $m$-step methods compute $\rho_{k+1}$ based on $m$ previous steps $\rho_{k}, \rho_{k-1}, \ldots, \rho_{k-m+1}$. Among these, the {\em Adams–Bashforth (AB)} methods form a class of explicit schemes with the general form
\begin{equation*}
\rho_{k+1} = \rho_k + h \left[ b_1 f(\rho_k) + b_2 f(\rho_{k-1}) + \cdots + b_m f(\rho_{k - m + 1}) \right],
\end{equation*}
given initial conditions $\rho_0, \rho_1, \ldots, \rho_{m-1}$. For example, the coefficients of the 2-step AB method are $b_1 = \frac{3}{2}$, $b_2 = -\frac{3}{2}$. Refer to \cite{Atkinson-et-al:2009} for more details and the list of higher order AB methods. 
AB methods give an accuracy of order $\mathcal{O}(h^m)$, but they typically have relatively small regions of absolute stability. To initialize the method, the starting values $\rho_1, \ldots, \rho_{m-1}$ can be computed using an RK method of global order $m-1$ (local order $m$).

The main advantage of an AB method of order $m$ over an RK method of the same order is a more computational efficiency in the sense that each AB step requires only one partial eigendecomposition, whereas an RK method of the same order requires $m$ such decompositions per step. However, RK methods have larger stability regions and usually give more accurate results for a fixed time step $h$.

For an efficient implementation, similar to \eqref{eq:rho-expand} we can write $\rho_{k+1}$ in terms of its low-rank components as
\begin{equation*}
\begin{split}
\rho_{k+1} &= \rho_k + h \left[ b_{1} f(\rho_{k}) + \cdots + b_m f(\rho_{k-m+1}) \right] \\
&= \Vt_k\Lt_0\Vt_k^* + h\left[ b_1 (Z_k\Vt_k^*+\Vt_kW_k^*) + \cdots + b_m (Z_{k-m+1}\Vt_{k-m+1}^*+\Vt_{k-m+1} W_{k-m+1}^*)
\right]\\
& =: A_k\Vt_k^*+ \Vt_k {B_k^\ast} + A_{k-1}\Vt_{k-1}^*+ \Vt_{k-1} {B_{k-1}^\ast} + \cdots + A_{k-m+1}\Vt_{k-m+1}^*+ \Vt_{k-m+1} {B_{k-m+1}^\ast}
\end{split}
\end{equation*}
where $\Vt_\ell$, $Z_\ell$, $W_\ell$ are the low-rank components derived from previous steps $\ell=k-m+1, \ldots, k$.
The same efficient strategy, outlined earlier for RK methods, is applied for storage, manipulation, and eigendecomposition in the AB schemes. 
These schemes are refereed to as {LREI-AB} methods.

\section{Solving q-LL equation}\label{sect:q-LL}

At first glance, one might expect solving the q-LL equation to be computationally cheaper than solving the q-LLG equation. However, if we manage to have numerical schemes that respect the physical invariants, and/or we aim to simulate for large spin systems (e.g., more than 13) by employing the low-rank representations, solving the q-LL equation exhibits roughly the same time and space complexities as the q-LLG equation. More precisely,
in the LREI process for the q-LL equation, we do not need to solve the Sylvester equation. The only difference then lies in the computation of the matrix blocks of $X$, which are now given by
\begin{equation}\label{eq:X-blocks-sol-qLL}  
\begin{split}
   & X_{11} = \frac{\ii}{\hslash} (\Lt\Vt^*H\Vt-\Vt^*H\Vt\Lt)-\frac{\kappa}{\hslash}(\Lt^2\Vt^*H\Vt+\Vt^*H\Vt\Lt^2)+\frac{2\kappa}{\hslash}\Lt\Vt^*H\Vt\Lt ,\\
   & X_{12} = \frac{1}{\hslash}(\ii I-\kappa \Lt)(\Lt\Vt^*H\Vh). 
\end{split}
\end{equation}
A comparison of \eqref{eq:X-blocks-sol} and \eqref{eq:X-blocks-sol-qLL} shows that the computational cost of computing $X_{11}$ and $X_{12}$ 
shares the same dominant matrix-matrix operations resulting in a complexity of the same order. In particular, solving the Sylvester equation in the q-LLG case does not introduce any additional cost for the LREI algorithm. All other components of the algorithms (including the implementation in RK and AB settings) are exactly the same for both the q-LL and q-LLG equations. 

\section{Numerical results}\label{sect:numerical}

The code was implemented in MATLAB and executed on a MacBook Pro with an Apple M1 Max chip and 64 GB of RAM. As a Krylov subspace method for computing partial eigendecompositions we use the \texttt{eigs} function of MATLAB which internally uses ARPACK, an iterative Arnoldi/Lanczos method \cite{Lehoucq-et-al:1998}. This function can also work with function handles such as $g$ in \eqref{eq:g-func}. 

 The spin system we consider here is governed by the spin Hamiltonian
\begin{equation}\label{eq:H1}
H = \frac{2J}{\hslash^2}\sum_{ij} \mathbf{S}_i \cdot \mathbf{S}_j + \frac{2}{\hslash^2}\sum_{ij} \bm{d}_{ij} \cdot \left[\mathbf{S}_i \times \mathbf{S}_j \right] -\mu \sum_i \bm{b}\cdot \mathbf{S}_i
\end{equation}
where $J\in\R$ is the isotropic Heisenberg exchange interaction, $\bm{d}_{ij}\in \R^3$ ($\bm{d}_{ij}=-\bm{d}_{ji}$)
represents the Dzyaloshinskii–Moriya interaction (DMI), and $\bm{b} \in \mathbb{R}^3$ is a uniform external magnetic field. The gyromagnetic constant is given by $\mu = -\frac{\mu_B g}{\hslash}$, where $\mu_B = 5.8 \times 10^{-2}\, \text{meV/T}$ is the Bohr magneton and $g = 2$ is the Landé $g$-factor.
The spin operators $\mathbf{S}_i = [S_i^x, S_i^y, S_i^z]$ are spin-$\frac{1}{2}$ operators defined using the Pauli matrices as described in \eqref{eq:spin_op}. An exact solution to the system is known when the initial density matrix $\rho_0$ has rank one; see \cite{azimi-mirzaei:2025}. In all examples, the fixed dissipation parameter $\kappa=0.5$ will be used. 

\subsection{Accuracy test}

First we reproduce the results of \cite{azimi-mirzaei:2025} using the LREI methods to verify their accuracy for a rank-one initial density matrix for which the exact solutions are available. We consider an antiferromagnetically ordered pure state
$\rho_0 = \ket{\text{AF}_{1}}\bra{\text{AF}_{1}}$
as such a rank-one density matrix\footnote{
The Dirac notation $\ket{\psi}$ denotes a column vector and $\bra{\psi}$ its conjugate transpose (row vector). Consequently, the outer product $\ket{\psi}\bra{\psi}$ is a rank-1 matrix. In particular, the state $\ket{\text{AF}_{1}}$ corresponds to a vector with a single nonzero entry at the index $i = ({0101}\ldots)_2 + 1,$
where the binary string of length $n$ alternates between $0$ and $1$.
}. 

\begin{figure}[ht!]
    \centering
    \includegraphics[width=0.4\linewidth]{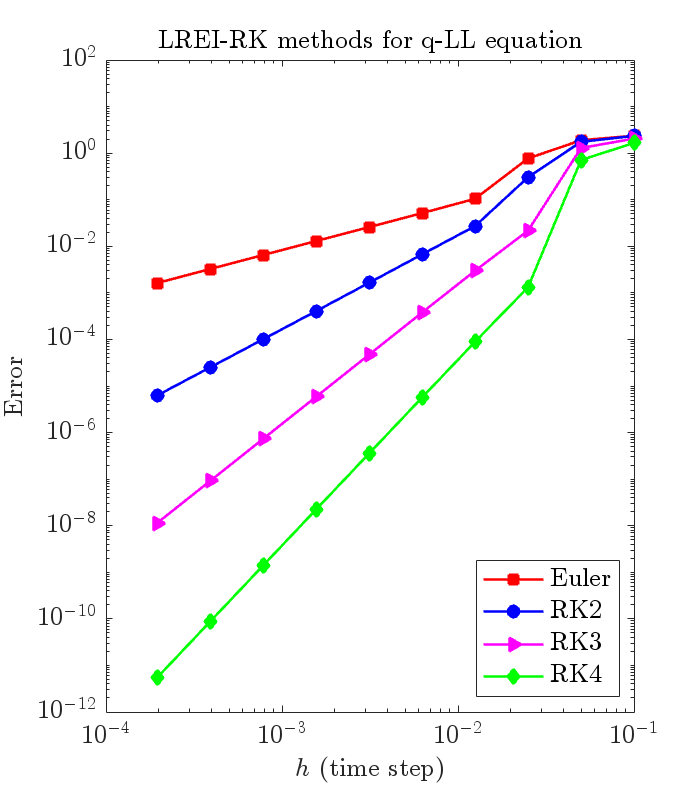}
    \includegraphics[width=0.4\linewidth]{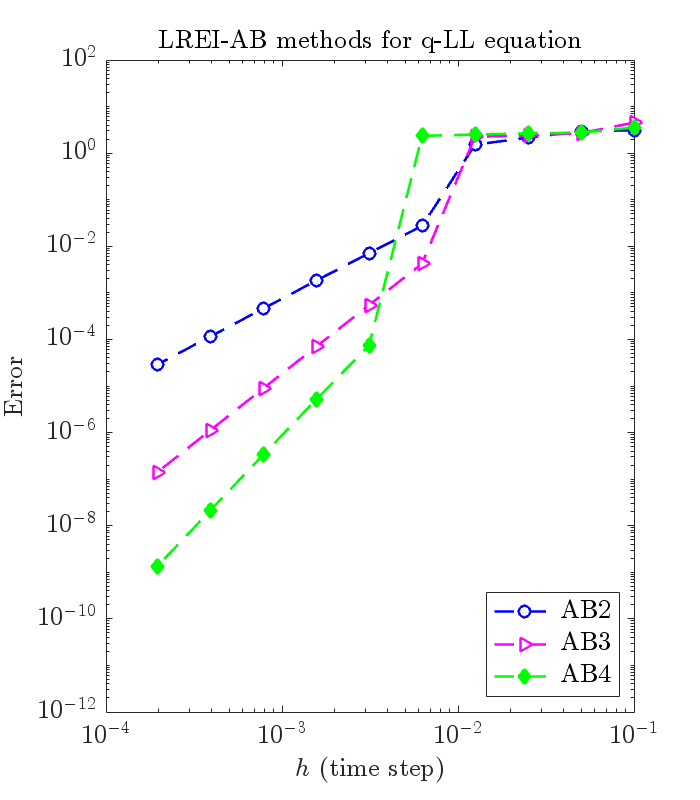} \\
    \includegraphics[width=0.4\linewidth]{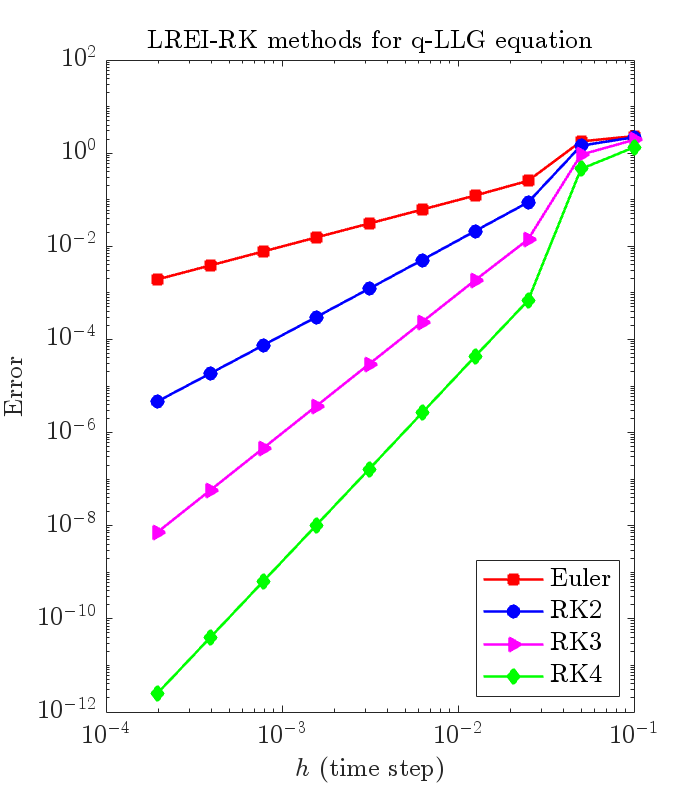}
    \includegraphics[width=0.4\linewidth]{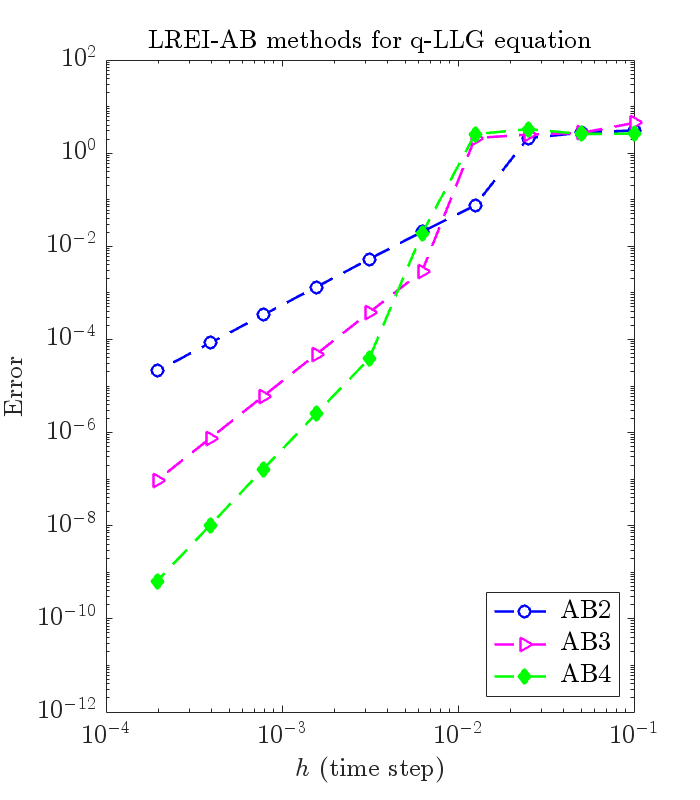} 

    \caption{Error plots of the conservative LREI-RK (left) and LREI-AB (right) methods for the 16-spin system (9 effective spins) modeled with q-LL (up) and q-LLG (down) equations. The theoretical orders $\mathcal O(h^m)$ are achieved for both RK$m$ and AB$m$ methods for sufficiently small values of time step sizes. However, RK methods are more accurate for a certain value of $h$.} 
    \label{fig:error_cmp}
\end{figure}

Following the setup in \cite{azimi-mirzaei:2025}, we consider a triangular lattice of 16 periodic spins (with 9 effective degrees of freedom) with Hamiltonian parameters $J = 1$ meV, $\bm{d}_{ij}=-\bm{d}_{ji}=\|\bm{d}\|(0, 0, 1)$, $\|\bm{d}\|_2=0.4$ meV, and $\bm{b} = \|\bm{b}\|_2(1,0,0)$, where $\|\bm{b}\|_2=1$ T is the strength of magnetic field vector. Given the pure state initial condition $\rho_0 =\ket{\text{AF}_{1}}\bra{\text{AF}_{1}}$, Figure \ref{fig:error_cmp} presents the error plots for LREI-RK and LREI-AB
methods in terms of the time step size $h$ for both q-LL and q-LLG equations.

As seen in the plots, RK methods achieve their expected convergence orders $m = 1, 2, 3, 4$ when $h \lesssim 0.05$. The AB methods also display their theoretical orders $m = 2, 3, 4$, but only for smaller step sizes, here $h \lesssim 0.0125$. This delay is attributed to the relatively smaller stability regions of high-order AB schemes.
For sufficiently small $h$, both RK$m$ and AB$m$ methods attain their theoretical convergence orders $\mathcal{O}(h^m)$. However, for the same order, RK methods tend to produce more accurate results than their AB counterparts, albeit at the cost of requiring multiple partial eigendecompositions per time step.

We also emphasize that the construction of LREI methods 
guarantees that both RK and AB schemes introduced in Sections \ref{sect:rk} and \ref{sect:ab} are conservative, in the sense that the spectrum of $\rho_k$ is preserved at each time step. As a result, physical properties such as the {\em non-negativity} of $\rho_k$ and the {\em trace} of all powers $\Tr(\rho_k^\ell),\, \ell=1,2,3,\ldots$ are preserved throughout the time evolution.

\begin{figure}[ht!]
    \centering
    \includegraphics[width=0.8\linewidth]{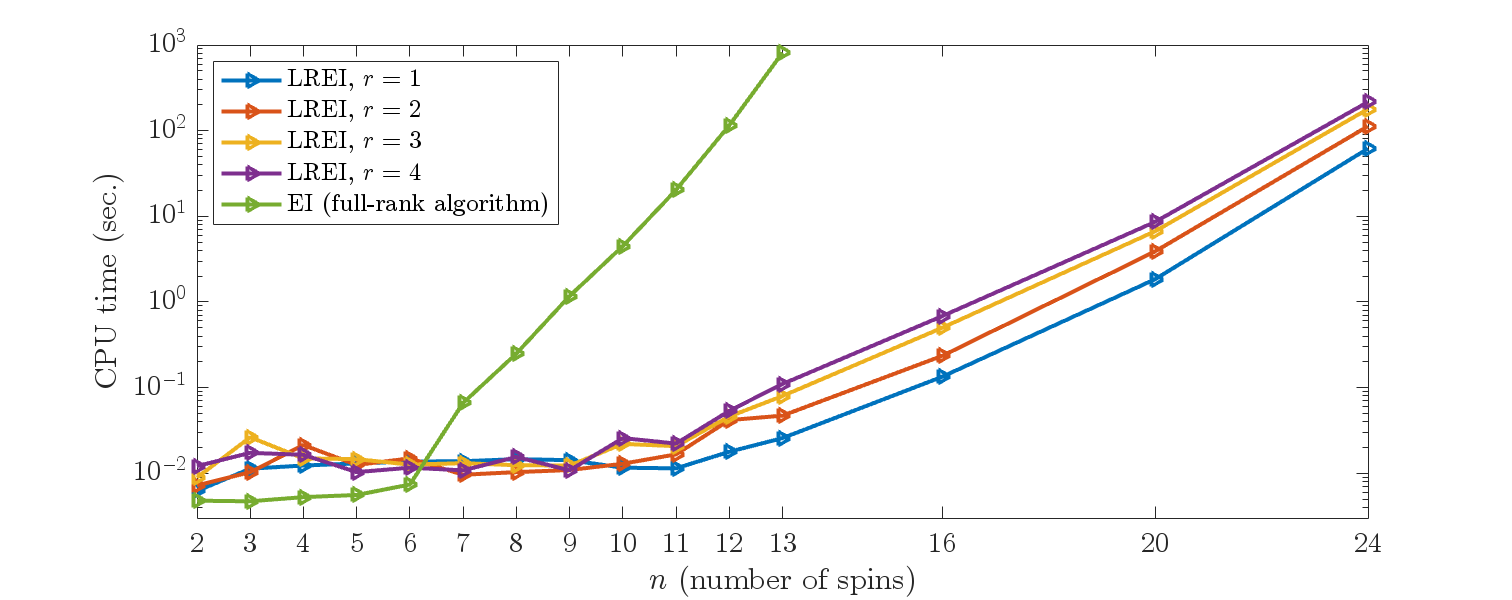}
    \caption{
Comparison of execution times for completing one step of the RK4 scheme using the original EI algorithm and the proposed LREI
method for q-LLG model. The LREI method is tested for initial density matrices of ranks $r=1,2,3,4$ while the EI method is tested only for a rank-1 initial data, as its performance is largely insensitive to matrix rank. The EI algorithm fails to run for systems with more than $13$ spins due to memory or computational limitations, whereas the LREI method successfully scales up to $n=24$.} 
\label{fig:time_cmp}
\end{figure}

\subsection{Complexity test}
We now compare the CPU time required by the original method presented in \cite{azimi-mirzaei:2025} (the EI method) and LREI method  
developed in this work. To evaluate performance, both algorithms are applied to spin clusters of different sizes, with spin number $n$ ranging from $2$ to $24$. In all cases, we use triangular lattice configurations (either periodic or non-periodic) and the Hamiltonian defined in \eqref{eq:H1}, with parameters identical to those used in the previous accuracy tests.

We focus on the RK4 and AB4, and perform a comparison based solely on execution times, since both low-rank and full-rank algorithms yield approximately the same accuracy in cases where the later one is computationally feasible. 
In Figure \ref{fig:time_cmp} we report the average CPU time required to complete `one' time step of the time evolution of q-LLG dynamics including computation of the density operator and the physical quantities such as energy, magnetization, and entanglement. For the new algorithm we report the results for initial conditions with different ranks ranging from $r = 1$ to $4$  while the original EI method is tested only for an initial data of rank $1$, as its performance is approximately independent of matrix rank. To construct initial conditions with different ranks, we use weighted combinations of rank-1 state
$\ket{\text{AF}_1}\bra{\text{AF}_1}$ (as defined above),
the antiferromagnetic state
$ \ket{\text{AF}_2}\bra{\text{AF}_2}$ with the index $i = ({1010...})_2 + 1$, the GHZ-State (Greenberger–Horne–Zeilinger) $ \ket{\text{GHZ}}\bra{\text{GHZ}}$, and the W-state $ \ket{\text{W}}\bra{\text{W}}$. See  \cite{azimi-mirzaei:2025} for definitions.

Figure \ref{fig:time_cmp} illustrates that while both algorithms execute in a fraction of a second for small systems (up to 5 spins), the low-rank approach demonstrates a clear and substantial computational advantage as the number of spins increases.
For instance, with $n = 13$ spins, the LREI
algorithm completes in less than one second, whereas the standard method requires approximately $10^4$ seconds. Moreover, the EI 
algorithm fails to execute for systems with more than $13$ effective spins due to memory-bound and computational constraints, while the LREI
method reliably scales up to $n = 24$ effective spins ($35$ periodic spins) on our personal laptop. Another notable observation is that increasing the matrix rank has only a modest and manageable impact on the computational complexity of the LREI algorithm. In practice, we observed linear growth in $r$, whereas the theoretical bound predicts a quadratic dependence.

\begin{figure}[ht!]
    \centering
    \includegraphics[width=0.8\linewidth]{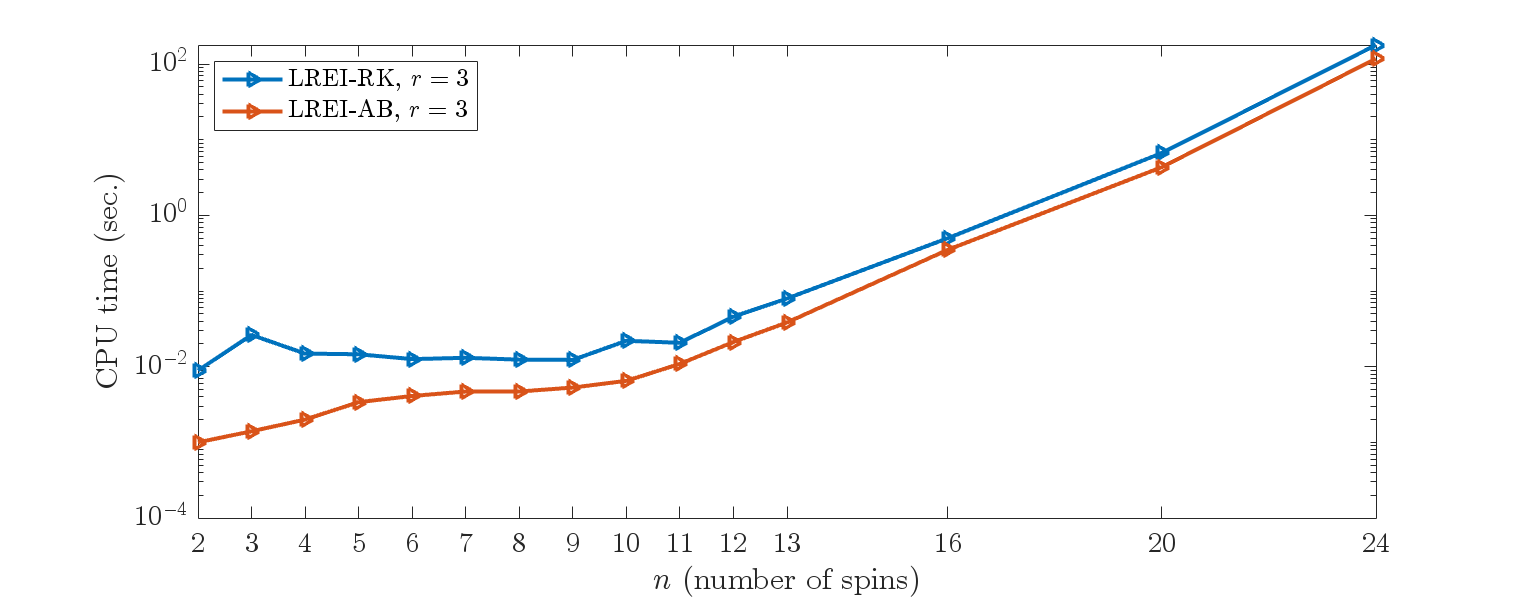}
    \caption{
Comparison of execution times for completing one step of the LREI-RK4 and LREI-AB4 schemes. Both algorithms are tested for an initial density matrix of rank $r=3$. For $n>10$, the LREI-RK method is, on average, twice as fast.}
\label{fig:time_cmp_ab4}
\end{figure}

Figure \ref{fig:time_cmp_ab4} compares the runtimes per time step of the LREI-RK4 and LREI-AB4 algorithms, tested on an initial density matrix of rank $r=3$. The LREI-AB4 method achieves an average speedup of about $2\times$ for $n > 10$. This is because LREI-RK4 requires four partial eigendecompositions per step, whereas LREI-AB4 needs only one. However, additional shared matrix–vector and matrix–matrix operations reduce the theoretical speedup from $4\times$ to $2\times$. This does not imply that LREI-AB schemes are preferable. As discussed earlier, LREI-RK methods generally achieve higher accuracy for a given step size $h$. Therefore, to reach comparable accuracy, the LREI-AB method typically requires a smaller step size, which diminishes its apparent efficiency advantage.

\subsection{q-LL vs. q-LLG dynamics}
As noted in the introduction, for a pure initial state $\rho_0$, the solutions of the q-LL and q-LLG equations coincide up to a time rescaling $t \mapsto t/(1+\kappa^2)$. However, no theoretical results are currently available for the case of mixed (non-pure) states. In this section, we compare numerical results for the q-LL and q-LLG equations in a system of nine effective spins, considering both pure and mixed initial states.

\begin{figure}[ht!]
    \centering
    \includegraphics[width=0.49\linewidth]{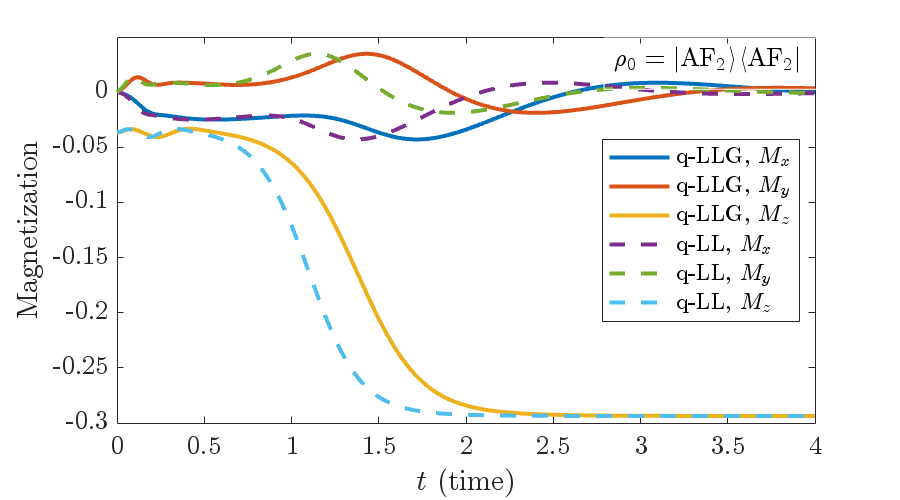}
    \includegraphics[width=0.49\linewidth]{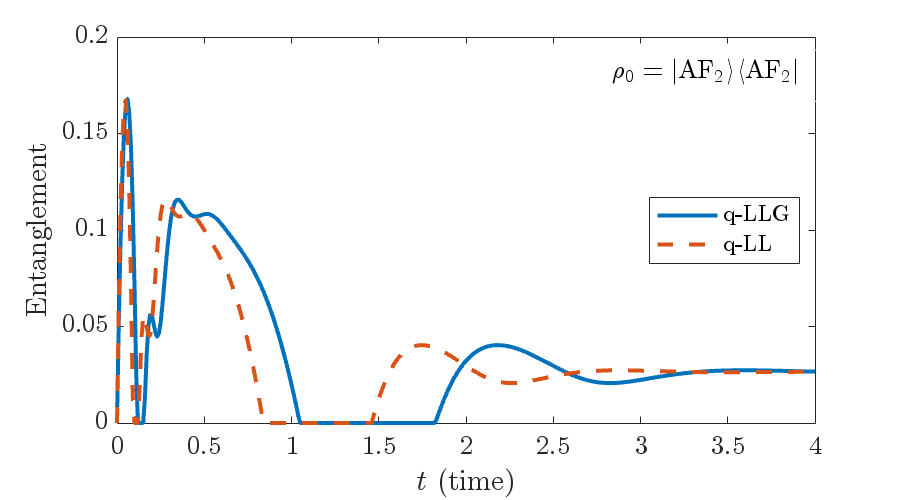}\\
    \includegraphics[width=0.49\linewidth]{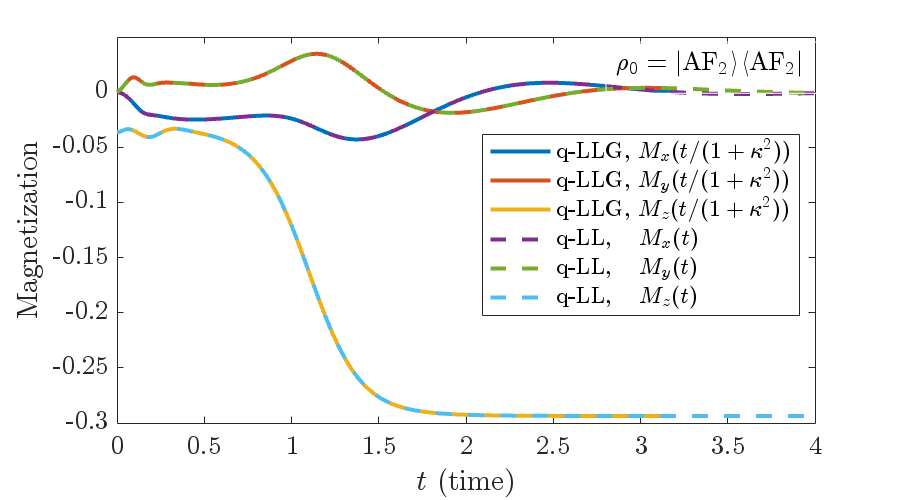}
    \includegraphics[width=0.49\linewidth]{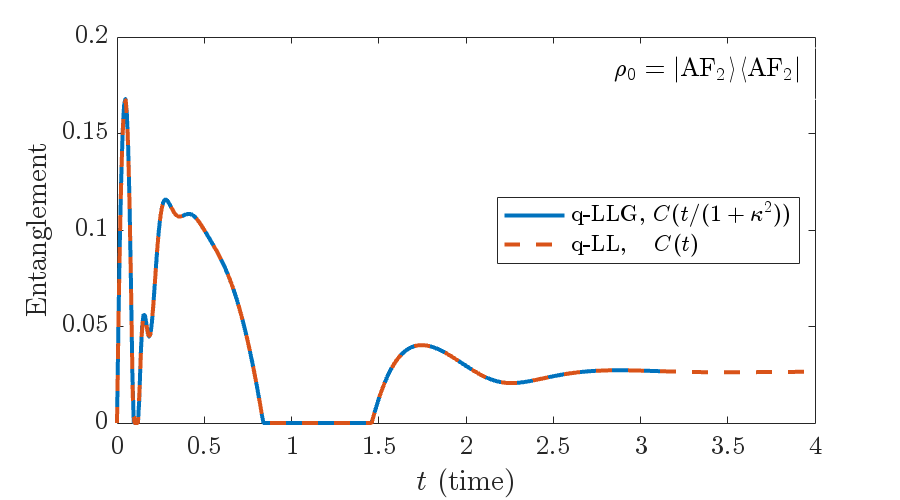}    
    \caption{
Comparison of q-LL and q-LLG dynamics of magnetization and entanglement (concurrence) for a pure initial state. In the first row, both q-LL and q-LLG solutions are plotted against $t$. In the second row, the q-LLG solutions are plotted against the rescaled time $t/(1+\kappa^2)$, demonstrating that for pure states the two dynamics coincide up to a time rescaling.
} 
\label{fig:cmp_pure}
\end{figure}

\begin{figure}[ht!]
    \centering
    \includegraphics[width=0.49\linewidth]{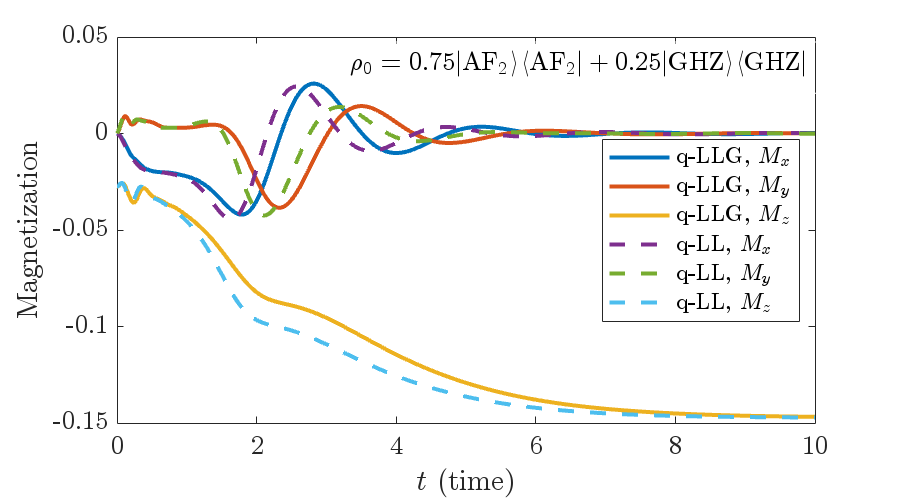}
    \includegraphics[width=0.49\linewidth]{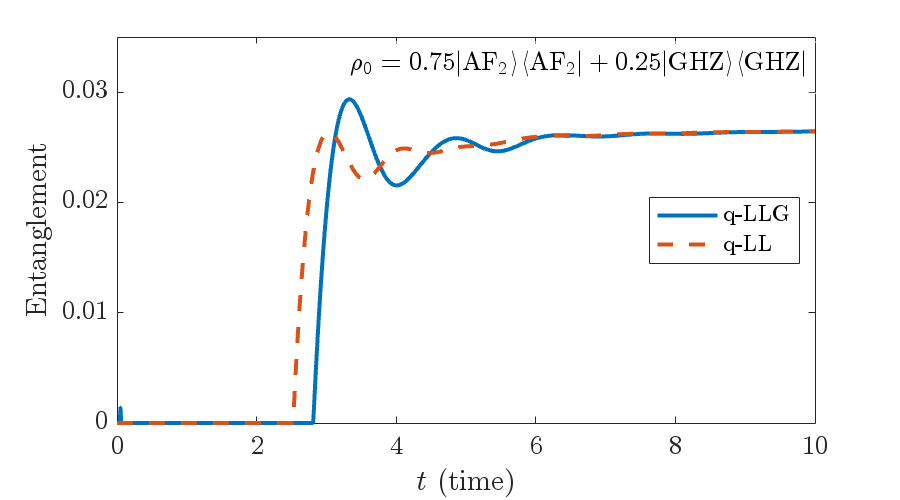}
    \caption{
Comparison of q-LL and q-LLG dynamics of magnetization and entanglement (concurrence) for a rank-2 (mixed) initial state. Both solutions are plotted against $t$. While the dynamics share certain similarities, they no longer coincide under the rescaled time $t/(1+\kappa^2)$.
} 
\label{fig:cmp_mixed}
\end{figure}

We compare the dynamics of magnetization and entanglement (concurrence) under the q-LL and q-LLG models with damping parameter fixed at $\kappa=0.5$. For the pure state case, as we observe from Figure \ref{fig:cmp_pure}, the system is initialized in the antiferromagnetic state $\rho_0 = \ket{\text{AF}_2}\bra{\text{AF}_2}$, and the two dynamics coincide up to a simple rescaling of time, $t \mapsto t/(1+\kappa^2)$. In the mixed state case, the initial density matrix is given by a convex combination $\rho_0 = 0.75 \ket{\text{AF}_2}\bra{\text{AF}_2} + 0.25 \ket{\text{GHZ}}\bra{\text{GHZ}}$. Figure \ref{fig:cmp_mixed} shows that in this case the q-LL and q-LLG dynamics again exhibit qualitative similarities but no longer coincide under the same time rescaling. This highlights the distinct behavior of the two models beyond the pure-state regime. 

Moreover, as expected since the q-LL model is only an approximation of the full q-LLG dynamics, our numerical experiments (not shown here) indicate that the solutions of the two models converge for smaller values of $\kappa$, while they diverge significantly for larger $\kappa$.

\subsection{A full-rank state}
We close this section with a remark about a well-known high-rank quantum state and how it can be efficiently handled using the LREI method.

The Werner state is a well-known class of mixed quantum states sometimes used as an initial condition in simulations. It is formed as a convex combination of a maximally entangled state and the maximally mixed state, and can be expressed as
$$
\rho_W = pI/2^n + (1-p)\hat \rho_0 , \quad p \in (0,1),
$$
where $\hat\rho_0$ is a low-rank maximally entangled state, $I$ is the identity matrix, and $I/2^n$ represents the maximally mixed state of $n$ qubits. The parameter $p$ controls the balance between the low-rank entangled part and the mixed noise contribution which allows the entanglement to be tuned continuously. This makes the Werner state an ideal test case for studying how the dynamics preserve or degrade quantum correlations.
Although $\rho_W$ is not low-rank (because the identity term $I/2^n$ is full-rank), it differs from a low-rank density matrix $\hat\rho_0$ only by a renormalization. Therefore, the LREI method can still be applied to solve the q-LL and q-LLG equations with $\rho_W$ as an initial condition. In fact, if $\rho(t)$ is the solution of either \eqref{eq:q-LL} or \eqref{eq:q-LLG} with initial condition $\rho_W$, we can write
\begin{equation}\label{eq:werner-sol}
    \rho(t) = p\,{I}/{2^n} + (1-p)\hat{\rho}(t),
\end{equation}
where $\hat{\rho}(t)$ is the solution of modified versions of \eqref{eq:q-LL} and \eqref{eq:q-LLG} where the damping parameter $\kappa$ is replaced by $(1-p)\kappa$, and with the low-rank initial state
$\hat\rho_0$.
We then apply the LREI method to approximate $\hat{\rho}(t)$ and finally  recover $\rho(t)$ via \eqref{eq:werner-sol}. 
The computation of physical quantities involving expressions of the form $\Tr(A \rho(t))$ for certain sparse observables $A$, simplifies to
$p/2^n\Tr(A) + (1-p)\Tr(A\hat\rho(t))$. The first term is cheap to evaluate due to the sparsity of $A$, while the second term can be efficiently computed using the method described in Subsection \ref{sect:rho-computation}.

\section{Summary and future studies}\label{sect:summury}

We presented a low-rank numerical algorithm, called low-rank eigenmode integration (LREI) method for solving quantum Landau–Lifshitz (q-LL) and quantum Landau–Lifshitz–Gilbert (q-LLG) equations using Runge-Kutta and Adams-Bashfoth ODE solvers.  The LREI method resulted in a significant improvement over the previous full-rank approach. By exploiting the low-rank structure of the density matrix and the sparsity of the operators involved, the proposed method significantly reduces both memory usage and computational cost, making it possible to simulate spin systems of considerably larger size. 
More precisely, the cost of each time-integration step drops from $\mathcal{O}(N^3)$ to $\mathcal{O}(r^2N)$, where $N = 2^n$ is the dimension for $n$ spins, and $r \ll N$ is the numerical rank of the solution matrix.
Likewise, the memory requirement is reduced from $\mathcal{O}(N^2)$ to $\mathcal{O}(rN)$, since we never form any full $N\times N$ (or nearly full) matrices during the computation.
Among several technical improvements we applied, one key idea was to handle the computation of the action of the invariant subspace of the density matrix associated with its zero eigenvalues. This was accomplished by applying Householder reflectors constructed for the {\em dominant} eigenspace, thereby enabling the entire solution process to proceed without ever forming any large matrices.

In addition, similar to the original method, the LREI 
method preserves physical properties such as spectrum preservation, trace conservation, and positivity of the density matrix in time evolution. 
  
In numerical experiments, we demonstrated that the LREI 
algorithm achieves the same level of accuracy as the original full-rank method, while scaling up efficiently to systems with up to 24 effective spins (35 periodic spins) on our personal laptop, which is well beyond the reach of the previous technique.
For larger spin systems, as low-rank representations and massive (sparse) Hamiltonian components still require storing and manipulating sizable matrices, the bottleneck is memory, not computation time or lack of parallelization. The simulation can be extended beyond 24 spins by adapting and executing the code on computers with higher memory capacity or HPC platforms either through nodes with higher RAM (fat nodes) or by using distributed memory architectures across nodes.

Here we highlight two avenues for future research: (i) developing adaptive time-stepping schemes that adjust the step size using local error estimates to improve efficiency without sacrificing accuracy; and (ii) advancing beyond standard Krylov-type partial eigensolvers by employing state-of-the-art randomized methods to further accelerate computation of the required eigenmodes for large spin systems.

From a quantum physics perspective, our new algorithm, LREI, makes it possible to study dynamical quantum phenomena in many-body spin systems that were previously out of reach. It gives researchers a new tool to uncover subtle quantum effects, gain a deeper understanding of the quantum properties and behavior of complex magnetic systems, and push forward the development of quantum technologies that use spin-based phenomena.


\bibliographystyle{plain}
\bibliography{dmref}

\end{document}